\documentclass[amsmath,amssymb,reprint,aps,prb,superscriptaddress]{revtex4-1}
\usepackage{graphicx}
\usepackage{dcolumn}
\usepackage{bm}
\usepackage[utf8]{inputenc}
\usepackage{braket}
\usepackage{hyperref}
\usepackage[mathlines]{lineno}
\usepackage{siunitx}
\usepackage{tabularx}
\usepackage{multirow}
\hypersetup{
    bookmarks=true,      
    unicode=false,          
    pdftoolbar=true,        
    pdfmenubar=true,        
    pdffitwindow=false, 
    pdfstartview={FitH},
    pdftitle={},   
    pdfauthor={Ruidy NEMAUSAT},     
    pdfsubject={Science},
    pdfcreator={Creator}, 
    pdfproducer={},
    pdfkeywords={phonon} {XANES} {NMR},
    pdfnewwindow=true,  
    colorlinks=true,      
    linkcolor=blue,     
    citecolor=blue,    
    filecolor=blue,
    urlcolor=blue}

\begin{document}

\title{Phonon effects on x-ray absorption and nuclear magnetic resonance spectroscopies}


\author{Ruidy Nemausat}
\email{ruidy.nemausat@impmc.upmc.fr}
\affiliation{Sorbonne Universit\'es, UPMC Univ Paris 06, Institut de Min\'eralogie, de Physique des Mat\'eriaux et de Cosmochimie (IMPMC), UMR CNRS 7590, 4 place Jussieu, F-75005, Paris, France}
\affiliation{Sorbonne Universit\'es,  UPMC Univ Paris 06, Laboratoire de Chimie de la Mati\`ere Condensée de Paris (LCMCP), UMR CNRS 7574,  Coll\`ege de France, 11 place Marcelin Berthelot, F-75005 Paris, France}

\author{Delphine Cabaret}
\affiliation{Sorbonne Universit\'es, UPMC Univ Paris 06, Institut de Min\'eralogie, de Physique des Mat\'eriaux et de Cosmochimie (IMPMC), UMR CNRS 7590, 4 place Jussieu, F-75005, Paris, France}

\author{Christel Gervais}
\affiliation{Sorbonne Universit\'es,  UPMC Univ Paris 06, Laboratoire de Chimie de la Mati\`ere Condensée de Paris (LCMCP), UMR CNRS 7574,  Coll\`ege de France, 11 place Marcelin Berthelot, F-75005 Paris, France}

\author{Christian Brouder}
\affiliation{Sorbonne Universit\'es, UPMC Univ Paris 06, Institut de Min\'eralogie, de Physique des Mat\'eriaux et de Cosmochimie (IMPMC), UMR CNRS 7590, 4 place Jussieu, F-75005, Paris, France}

\author{Nicolas Trcera}
\affiliation{Synchrotron SOLEIL, L'Orme des Merisiers,  St Aubin, BP 48, F-91192 Gif sur Yvette, France}
 
 \author{Am\'elie Bordage}
\affiliation{Universit\'e Paris-Sud, Institut de Chimie Mol\'eculaire et des Mat\'eriaux d'Orsay (ICMMO), UMR CNRS 8182, 15 rue du doyen Georges Poitou, F-91400, Orsay, France. }

\author{Ion Errea}
\affiliation{Fisika Aplikatua 1 Saila, EUITI, University of the Basque Country (UPV/EHU), Rafael Moreno ``Pitxitxi" Pasealekua 3, 48013 Bilbao, Basque Country, Spain}
\affiliation{Donostia International Physics Center (DIPC), Manuel Lardizabal Pasealekua 4, 20018 Donostia-San Sebasti\'an, Basque Country, Spain }

 \author{Francesco Mauri}
\affiliation{Sorbonne Universit\'es, UPMC Univ Paris 06, Institut de Min\'eralogie, de Physique des Mat\'eriaux et de Cosmochimie (IMPMC), UMR CNRS 7590, 4 place Jussieu, F-75005, Paris, France}

\date{\today}


\begin{abstract}

In material sciences, spectroscopic approaches combining \textit{ab initio} calculations with 
experiments are commonly used to accurately analyze the experimental spectral data. 
Most state-of-the-art first-principle calculations are usually performed assuming an equilibrium 
static lattice. Yet, nuclear motion affects spectra even when reduced  to the zero-point motion at
 0~K.  We propose a framework based on Density-Functional Theory   that includes quantum 
 thermal fluctuations in theoretical X-ray Absorption Near-Edge Structure (XANES) and solid-state 
 Nuclear Magnetic Resonance (NMR) spectroscopies and allows to well describe temperature effects 
 observed  experimentally. Within the Born-Oppenheimer and quasi-harmonic approximations, we 
  incorporate the nuclear motion by generating several non-equilibrium configurations from the dynamical 
  matrix. The averaged calculated XANES and NMR spectral data have been compared to experiments 
  in MgO,  proof-of-principle compound. The good agreement obtained between experiments and 
  calculations validates the developed approach, which suggests that calculating the XANES spectra
   at finite temperature by averaging individual non-equilibrium configurations is a suitable approximation.
    This study  highlights the relevance of phonon renormalization and the relative contributions of
     thermal expansion and nuclear dynamics on NMR and XANES spectra on a wide range of temperatures.

\end{abstract}


\maketitle

\section{Introduction}
\label{sec:intro}

X-ray Absorption Near-Edge Structure (XANES)\cite{Henderson2014} and solid-state Nuclear 
Magnetic Resonance (NMR)\cite{MacKenzie2002}  spectroscopies are powerful probes of the
 electronic and local structure of inorganic materials. The combination of these two techniques 
 provides a deep understanding of the electronic and structural properties of materials. 
 For instance, a recent study coupling X-ray absorption spectroscopy and NMR successfully
  resolved the local structure of Al sites in zeolites.\cite{vjunov2014} The noticeable improvements
   in methodology and instrumentation resulted in a spectacular enhancement of the quality and
    resolution of spectra for both techniques.\cite{Ashbrook2014,Salager2009,Milne2014,Koide2014} 
    Nonetheless, the huge amount of information contained in experimental spectra makes their 
     accurate assignment difficult. To address this difficulty theoretical tools are used beforehand
      or in conjunction with experimental data.\cite{Rehr2005,Charpentier2011,Bonhomme2012} 

Most first-principle calculations in the solid-state consider the nuclei fixed at their equilibrium 
positions, as obtained by X-ray or neutron diffraction. However, atoms are subjected to 
quantum thermal fluctuations, which reduce to the zero-point motion at the absolute
zero (0~K).  Temperature-dependent experiments exhibit significant variations of  
spectroscopic properties, such as  chemical shifts,\cite{Fiske1994,Webber2010}  relaxation 
times,\cite{Chan-Huot2015} and  Electric Field Gradient
 values \cite{Ostafin2007,Bonhomme2010,Braunling2010,ODell2011} (EFG) measured by 
 NMR or the pre-edge structures observed in XANES spectra of oxide 
 materials.\cite{Nozawa2005,Durmeyer2010,Manuel2012} In the case of NMR, it has been 
 observed that the chemical shifts vary of several ppm when  temperature increases over 
  thousands of degrees.\cite{Fiske1994,Webber2010} The EFG tensor is also very sensitive 
  to temperature and nuclear quadrupole resonance experiments  have shown a variation of
   the resonances frequencies $\nu_{\pm,0}$ up to $-0.2$~kHz.K$^{-1}$.\cite{Ostafin2007,Bonhomme2010} 
   In the case of XANES spectroscopy, it  has been recently shown that the intensity and position
    of the pre-edge peak highly depend on temperature.\cite{Nozawa2005,Durmeyer2010,Manuel2012} 
    It has been demonstrated that the pre-edge structure is due to a violation of symmetry 
    induced by the quantum thermal fluctuations. For instance, at the Al $K$-edge of various 
    oxides, the nuclear motion is responsible for the appearance of a pre-edge peak 
    corresponding to the $1s \rightarrow 3s$ forbidden electronic transitions.\cite{Cabaret2009,Brouder2010,Manuel2012}

Multiple attempts to theoretically reproduce the lattice dynamical effects in solid-state  
spectroscopies have emerged in the literature. Pioneering approaches based on  
averaging  the chemical shielding tensors over different orientations of mobile species 
were proposed to include nuclear motion in  NMR calculations.\cite{Gervais2009,Folliet2011} 
In parallel, it was shown that small displacements of either the absorbing 
atom\cite{Ankudinov2005a,Cabaret2009,Manuel2012} or  the $1s$ initial wave function in 
the crude Born-Oppenheimer approximation\cite{Brouder2010}  could reveal forbidden 
transitions in $K$-edge XANES spectra.  Although quite promising, these methods do 
not account for the collective lattice dynamics. A substantial  theoretical  work  proved 
that vibrations could be represented as  the convolution of  the  x-ray absorption cross 
section, calculated for the equilibrium configuration, with the phonon spectral 
function.\cite{Fujikawa1999,Fujikawa2014} This theory has been  applied to reproduce 
room-temperature  experimental XANES spectra in Ref.~\onlinecite{Manuel2012}.  In 
the case of non-quantum nuclear motion, the method of choice would be Molecular 
Dynamics (MD) at finite temperature, either classical or \textit{ab initio}. 
In NMR, \textit{ab initio} MD was used to study the dependence on temperature of 
the chemical shift 
\cite{Dumez2009,Gortari2010,Webber2010,Robinson2010,Dracinsky2013}  and 
quadrupolar relaxation rates\cite{Carof2014,Chan-Huot2015} but mostly in organic 
compounds. MD calculations were also used in XANES to calculate  Li and S $K$-edges 
XANES spectra of Li-ion batteries at room temperature\cite{Pascal2014,Pascal2015} 
or the Al and Fe $K$-edges in warm dense plasmas.\cite{Peyrusse2010,Mazevet2014,Dorchies2015} 
However, MD consider the vibrations as a classical phenomenon and is therefore  
appropriate  only if $k_BT > \hbar\varpi_{vib}$, with $\varpi_{vib}$ the vibration 
frequency.\cite{Kuhne2007} To account for the quantum behavior of  vibrations, Path 
Integral Molecular Dynamics (PIMD) was  used to simulate NMR\cite{Dracinsky2014} 
and C $K$-edge NEXAFS\cite{Schwartz2009} spectra  in organic compounds. In both 
spectroscopies, the results were improved using PIMD, but at the cost of a larger statistical 
error. Seeking for computationally less expensive  methods, Monte Carlo sampling has 
been used to account for the vibrational effects on the chemical shifts in MgO\cite{Rossano2005} 
and recently a more computationally efficient method arose by assuming a quadratic coupling
 between vibrations and the shielding tensor.\cite{Monserrat2014}   The Monte Carlo sampling 
 method was also used to simulate XANES spectra of molecules in solutions,\cite{Canche-Tello2014} 
 but, to our knowledge, it has not been applied  to inorganic solids yet.

The purpose of this work is to describe quantum thermal fluctuations, using methods based
 on Density-Functional Theory (DFT).\cite{Jones2015} In the Born-Oppenheimer  
 (BO)\cite{Born1954} and Quasi-Harmonic Approximation (QHA),\cite{Baroni2010,Fultz2010} 
 the thermal effects are modeled by generating atomic position configurations obeying 
 quantum statistics at finite temperatures. The theoretical work is confronted to NMR and 
 recently acquired XANES temperature-dependent experiments for our proof-of-principle 
 compound MgO. This ionic oxide has been chosen for three main reasons: (i) its rock salt 
 structure is ideal to observe thermal effects as its lattice constant is the only  free parameter,
  (ii) it shows no phase transition up to its melting point around 3250~K,\cite{Ronchi2001} 
  allowing a wide temperature range for experimental measurements, (iii)  multiple theoretical
   studies demonstrated that the harmonic behavior of MgO remains at temperature as high 
   as 1500~K and its Debye temperature has been found, theoretically, to be about 
   941~K.\cite{Fincham1994,Gavartin2001,Oganov2003} 

The paper is structured as follows. In Sec.~\ref{sec:theory}, the formalism upon which our 
theoretical model is based is detailed. In Sec.~\ref{sec:expt}, experimental and computational 
details are given. In Sec.~\ref{sec:results},  experimental and theoretical results obtained 
on MgO are presented and discussed. Finally, the  conclusions of this work are drawn  
in Sec.~\ref{sec:end}.

\section{Formalism}
\label{sec:theory}

\subsection{Quasi-harmonic vibrations}

A crystal can be seen as a system of $N$ nuclei and $N_e$ electrons with respective position 
vectors ($\mathbf{R}_1, \dots, \mathbf{R}_{N}$) and ($\mathbf{r}_1, \dots, \mathbf{r}_{N_{e}}$). The collective
 coordinates $\mathbf{\overline{R}}=(\mathbf{R}_1, \dots, \mathbf{R}_{N})$ and $\mathbf{\overline{r}} = (\mathbf{r}_1, \dots, \mathbf{r}_{N_{e}})$ 
 are used thereafter. The stationary states of the system are described by the  wave function $\Psi(\mathbf{\overline{r},\overline{R}})$  
 solution of the following general Schr\"odinger equation,
\begin{equation}
\label{eq:hamtot} \left( T_N + T_e + V_e + V_{e-N} + V_{N}\right)  \Psi(\mathbf{\overline{r},\overline{R}}) = E\Psi(\mathbf{\overline{r},\overline{R}}),
\end{equation}
with $T_{N}$ the kinetic nuclear operator, $T_{e}$ the kinetic electronic operator, $V_{e}$ the Coulomb
 potential between electrons, $V_{e-N}$ the Coulomb potential between nuclei and electrons, and 
 $V_{N}$ the Coulomb potential between nuclei. The total energy of the crystal is denoted by $E$.

In the BO approximation, which assumes that the electronic cloud reacts instantaneously to the 
nuclear motion, the wave function solution of Eq.~(\ref{eq:hamtot}) can be approximated as the product
\begin{equation}
\label{eq:bowfc}\Psi_n^j(\mathbf{\overline{r},\overline{R}}) = \chi_n^j(\mathbf{\overline{R}})\psi_n({\mathbf{\overline{r};\overline{R}}}), 
\end{equation}
of an electronic part $\psi_n(\mathbf{\overline{r};\overline{R}})$, in which  $\mathbf{\overline{R}}$ is 
a parameter, and a nuclear part  $\chi_n^j({\mathbf{\overline{R}}})$. The  electron orbital index is $n$, 
 and $j$ indexes vibrational states. 

The Hamiltonian acting onto the electronic variables is labelled $H_{BO}$, 
\begin{equation}
H_{BO} = T_e + V_e + V_{e-N} + V_{N}
\end{equation} 
where $V_{N}$ is then a constant energy term determined for a  given nuclei configuration $\mathbf{\overline{R}}$. Thus $H_{BO}$ is parametrized by $\mathbf{\overline{R}}$. The electronic wave function $\psi_n(\mathbf{\overline{r};\overline{R}})$ is solution of the  Schr\"odinger  equation,
\begin{equation}
\label{eq:scropsi}  H_{BO} \psi_n(\mathbf{\overline{r};\overline{R}}) = \varepsilon_n(\mathbf{\overline{R}})\psi_n(\mathbf{\overline{r};\overline{R}}), 
\end{equation}
which introduces the energy surface $\varepsilon_n(\mathbf{\overline{R}})$. In the BO approximation
 the lattice dynamics is described by  
\begin{equation}
\label{eq:scrochi} \left[T_N + \varepsilon_{n}(\mathbf{\overline{R}})\right]\chi_n^j({\mathbf{\overline{R}}}) = E_n^j\chi_n^j({\mathbf{\overline{R}}})
\end{equation}
where  the vibrational wave functions $\chi_n^j(\mathbf{\overline{R}})$ are the orthonormal solutions 
and $E_n^j$ is the total energy of the whole crystal. Equations~(\ref{eq:scropsi}) and (\ref{eq:scrochi}) 
imply that $\{\psi_n\}$ and $\{\chi_n\}$ are complete basis sets of eigenvectors of $H_{BO}$ and of the
 nuclear Hamiltonian $\left[T_N + \varepsilon_{n}(\mathbf{\overline{R}})\right]$, respectively, leading to
  the completeness relations
\begin{eqnarray}
\sum_{n}\psi_n^*(\mathbf{\overline{r},\overline{R}})\psi_n(\mathbf{\overline{r}',\overline{R}}) &=& \delta(\mathbf{\overline{r}-\overline{r}}') ; \forall \mathbf{\overline{R}},\\
\sum_{j}\chi_n^j(\mathbf{\overline{R}})\chi_n^j(\mathbf{\overline{R}}') &=& \delta(\mathbf{\overline{R} - \overline{R}}') ; \forall n. 
\label{eq.chi}
\end{eqnarray}

The phonon-induced displacement of nucleus $I$ in the Cartesian direction $\alpha$ 
 \begin{equation}
\label{eq:defu} u_I^\alpha=R_I^\alpha - R_{I,\mathrm{(eq)} }^\alpha 
\end{equation}
is  small compared to the atomic bond length. Thus a Taylor expansion of the BO energy surface as a 
function of the nuclear displacements can be carried out and is truncated at the second order in the 
harmonic approximation. The energy scale is shifted such as the zero-order term is null and the first-order
 term vanishes because forces acting on individual nuclei are zero at equilibrium. In this approximation, 
 $\left[T_N + \varepsilon_{n}(\mathbf{\overline{R}})\right]$ becomes the nuclear harmonic Hamiltonian as
\begin{equation}
\label{eq:hamilton}\mathcal{H} = \sum_{I=1}^{N}\sum_{\alpha=1}^3\frac{\left(P_I^{\alpha}\right)^2}{2M_I} + \frac{1}{2}  \sum_{I,J}^{N} \sum_{\alpha,\alpha'}^3 u_{I}^\alpha \mathcal{C}_{\alpha\alpha'}^{IJ}  u_{J}^{\alpha'},
\end{equation}
where $T_N$ is explictly written in terms of  $\mathbf{P}_I$,  the momentum operator of the $I$th nucleus. 
In Eq.~(\ref{eq:hamilton}) we have  introduced the interatomic force constant matrix $\mathcal{C}$ whose elements are 
\begin{equation}
\mathcal{C}_{\alpha\alpha'}^{IJ} =   \left. \frac{\partial^2 \varepsilon_n(\mathbf{\overline{R}})}{\partial u_I^{\alpha}\partial u_J^{\alpha'}}\right|_{\mathrm{(eq)}}.
\end{equation}
Rescaling $\mathcal{C}$ by the nuclei masses define the  dynamical matrix, whose   diagonalization as 
\begin{equation}
\label{eq:diagmatdyn}\sum_{J=1}^{N}\sum_{\alpha'=1}^3 \frac{\mathcal{C}_{\alpha\alpha'}^{IJ}}{\sqrt{M_IM_J}}\epsilon_{J\mu}^{\alpha'} = \varpi^2_{\mu}\epsilon_{I\mu}^{\alpha},
\end{equation}
provides phonon polarization vectors $\epsilon_{I\mu}^{\alpha}$ and phonon frequencies $\varpi_{\mu}$,
 where $\mu$ indexes phonon modes. Using in Eq.~(\ref{eq:hamilton}) 
\begin{eqnarray}
\label{eq:norm1}u_I^{\alpha} &=& \sum_{\mu=1}^{3N} \frac{1}{\sqrt{M_I}} \epsilon_{I\mu}^{\alpha} q_{\mu},\\
\label{eq:norm2}P^{\alpha}_I &=& \sum_{\mu=1}^{3N} \sqrt{M_I} \epsilon_{I\mu}^{\alpha} p_{\mu},
\end{eqnarray}
which  introduce the normal coordinates $q_{\mu}$ and $p_{\mu} = \dot{q}_\mu$, $\mathcal{H}$ can 
be written as a sum of $3N$ independent Hamiltonian operators of harmonic  oscillators
\begin{equation}
\mathcal{H} = \sum_{\mu=1}^{3N} \frac{1}{2}\left( p_{\mu}^2 + \varpi_{\mu}^2q_{\mu}^2 \right).
\end{equation}

Nevertheless, a harmonic model considers the phonon normal modes as independent quasiparticles 
and does not account for any anharmonic phenomenon, such as  thermal expansion. In this work   
QHA is used to include thermal expansion effects. The model is no longer purely harmonic but does 
not describe phonon-phonon interaction, as the  phonon normal modes are still independent. Within  
QHA, the probability $\mathcal{P}(\overline{q}_\mu)$ of finding the system in any set of normal coordinates $\overline{q}_\mu$, is expressed as a product of Gaussian functions following a normal distribution whose widths depend on temperature and phonon frequency.\cite{Maradudin1963} The Gaussian functions are centered on  $q_\mu=0$, i.e., at the equilibrium position when $\mathbf{u}_I=\mathbf{0}$  (Eq.~\ref{eq:norm1}). 
The $\mathcal{P}(\mathbf{\overline{R}})$  probability distribution  is written  as
\begin{equation}
\label{eq:probaR}\mathcal{P}(\mathbf{\overline{R}}) = \mathcal{A}\exp \left( - \sum_{IJ\alpha\alpha'\mu} \frac{\sqrt{M_IM_J}}{2a_\mu^2} \epsilon_{I\mu}^\alpha \epsilon_{J\mu}^{\alpha'} u_I^\alpha u_J^{\alpha'}\right)
\end{equation}
with $\mathcal{A}$ a normalization constant. The $a_\mu$ normal length of the $\mu$ vibration 
mode  is the standard deviation of  collective normal coordinates $\mathbf{\overline{q}}$ and 
depends explicitly on  temperature $T$ and phonon frequencies $\varpi_\mu$
\begin{equation}
\label{eq:normlen}a_\mu = \sqrt{\frac{\hbar}{2\varpi_\mu}    \coth\left( \frac{\beta \hbar\varpi_\mu}{2}\right)}, 
\end{equation}
with $\beta=1/k_BT$. Equation~(\ref{eq:normlen}) describes how the phonon normal modes are
 thermally populated. The statistical average of any observable  $\mathcal{O}$ is then
\begin{equation}
\label{eq:statav}\left< \mathcal{O}\right> = \int d\mathbf{\overline{R}}\ \mathcal{O}(\mathbf{\overline{R}})\ \mathcal{P}(\mathbf{\overline{R}}).
\end{equation}

For a given crystal, after calculating and diagonalizing the dynamical matrix  (Eq.~\ref{eq:diagmatdyn})
 to obtain the phonon frequencies $\varpi_\mu$ and polarization vectors $\epsilon_\mu$,   a set of 
 $N_c$ nuclear configurations obeying the $\mathcal{P}(\mathbf{\overline{R}})$ quantum statistical 
 distribution  is generated. For each $\mu$ mode, a set $\{x_\mu^i\}_{i=1\dots N_c}$ of $N_c$ random 
 Gaussian numbers is created. Then, each $x_\mu$ is multiplied by the corresponding normal length 
 $a_\mu$, as defined in Eq.~(\ref{eq:normlen}).  The set of so-generated normal coordinates
  \{$q_\mu^i = x_\mu^i a_\mu\}_{i=1\dots N_c}$ obeys the probability distribution (Eq.~\ref{eq:probaR}). 
  The nuclear position collective vectors \{$\overline{\mathbf{R}}^i\}_{
i=1\dots N_c}$  are obtained using Eqs.~(\ref{eq:defu})~and~(\ref{eq:norm1}) such as
\begin{equation}
R_I^{\alpha, i} = R_{I,\mathrm{(eq)}}^{\alpha, i} + \sum_{\mu=1}^{3N}\frac{1}{\sqrt{M_I}}\epsilon_{I\mu}^{\alpha}a_\mu x_\mu^i. 
\end{equation}
According to the importance sampling technique, Eq.~(\ref{eq:statav}) is equivalent to
\begin{equation}
\label{eq:statavmc}\left< \mathcal{O}\right> \simeq \frac{1}{N_c} \sum_{i=1}^{N_c}  \mathcal{O}(\mathbf{\overline{R}}^i).
\end{equation}
In this work, $\mathcal{O}(\mathbf{\overline{R}}^i)$ is either  the nuclear magnetic shielding tensor, 
the EFG tensor or the XANES cross section for the $i$th configuration.

\subsection{Nuclear Magnetic Resonance spectroscopy}

To compare with NMR experiments, for each nucleus, the isotropic value of the magnetic shielding
 tensor $\boldsymbol\sigma$ is evaluated as
\begin{equation}
\sigma_{\mathrm{iso}} = \frac{1}{3}\mathrm{tr}\left(\boldsymbol\sigma\right).
\end{equation}
More precisely,  the isotropic chemical shift  $\delta_{\mathrm{iso}} \simeq -\left( \sigma_{\mathrm{iso}}
 - \sigma_{\mathrm{ref}}\right)$ is considered, where $\sigma_{\mathrm{ref}}$ is the isotropic 
 shielding of the standard reference. Unlike the Larmor frequency,   $\delta_{\mathrm{iso}}$ contains
  the response of the electronic system to the magnetic perturbation and  is therefore an explicit function
   of the electronic environment around the probed nucleus. Thus, $\delta_{\mathrm{iso}} = \delta_{\mathrm{iso}}(\mathbf{\overline{R}})$
    and $\braket{\delta_{\mathrm{iso}}(\mathbf{\overline{R}})}$ is obtained as in Eq.~(\ref{eq:statavmc}).

\subsection{XANES spectroscopy}

In a single-electron approach,  at the $K$-edge, the  XANES  cross-section, in the electric dipole 
approximation  is usually given by\cite{Brouder1990}
\begin{equation}
\label{eqxas} 	\sigma(\hbar\omega) = 4\pi^2\alpha_0\hbar\omega\sum_n \left| \bra{\psi_n} 
\mathbf{\hat{e}\cdot r}  \ket{\psi_{1s}} \right|^2 \delta({\varepsilon_n - \varepsilon_{1s} - \hbar\omega}).
\end{equation}
where $\hbar\omega $  and $\mathbf{\hat{e}}$ are the energy and polarization vector of the incident 
X-ray photon, respectively, and $\alpha_0$  the fine structure constant. The final $\ket{\psi_{n}}$   
and initial $\ket{\psi_{1s}}$ electronic states of  energy  $\varepsilon_{n}$ and $\varepsilon_{1s}$, 
respectively, are solution of Eq.~(\ref{eq:scropsi}). 

Equation~(\ref{eqxas}) rests upon  the assumption that the nuclei are fixed in a generic configuration,
 hence $\sigma$ is parametrized by $\mathbf{\overline{R}}$, i.e., $\sigma(\hbar\omega)=\sigma(\hbar\omega;\mathbf{\overline{R}})$.
  A more rigorous expression of the cross section accounting for the total system of nuclei and electrons is
\begin{equation}
\sigma_{tot}(\hbar\omega) = 4\pi^2\alpha_0 \hbar\omega \sum_{n,j} \left| \bra{\Psi_n^j} \mathbf{\hat{e}
 \cdot \overline{r}} \ket{\Psi_{1s}^{0}}\right|^2\delta(E_n^j - E_{1s}^{0} - \hbar\omega)
\label{eqxas2}
\end{equation}
where the  $\Psi_{1s}^{0}$ initial state and $\Psi_{n}^{j}$ final states are defined as in Eq.~(\ref{eq:bowfc}).
In Appendix~\ref{green}, it is demonstrated that $\sigma_{tot}$ can be expressed, using the first order of the
 expansion given in Eq.~(\ref{gtg}), by 
\begin{equation}
\label{eq:avt} \sigma_{tot}(\hbar\omega) =   \int d\mathbf{\overline{R}}\    \rho( \mathbf{\overline{R}})   
 \sigma(\hbar\omega;\mathbf{\overline{R}}),
\end{equation}
with $\rho( \mathbf{\overline{R}})$ the quasi-harmonic weighting displacement distribution. The next orders of the 
expansion in Eq.~(\ref{gtg})  are not considered in the present work.  In Eq.~(\ref{eq:avt}), $\sigma_{tot}$  is  the 
average of  individual configuration cross sections $\sigma(\hbar\omega;\mathbf{\overline{R}})$ using a   
probability distribution that takes into account the temperature and vibrations frequencies in a  form consistent 
with  Eq.~(\ref{eq:statavmc}).

\section{Experimental and calculation details}
\label{sec:expt}

\subsection{Experimental setup}

Mg $K$-edge X-ray absorption experiments were performed at the LUCIA beam line (SOLEIL Saint-Aubin, France).\cite{Lucia06} 
The incident energy range (1280--1400~eV) was selected to include the Mg $K$-edge using a double Beryl  monochromator. 
The pressure in the experimental chamber was  10$^{-7}$ mbar. The incident X-ray beam was set to a spot size of 
$1\times2$~\si{\milli\metre^2}. Temperature-dependent measurement were conducted at 300~K, 573~K, 773~K, 873~K
 and 1273~K using a boron nitride furnace. Only the spectra recorded at 300~K, 573~K and 873~K are shown. 
 The 4~cm$^2$ MgO single crystal was held using a perforated lamella of molybdenum. The  temperature of the
  sample was measured using a Chromel-Alumel thermocouple. The spectra were recorded in fluorescence mode
   with a four element Silicon Drift Diode detector, protected from infrared and visible radiations by a thin beryllium 
   window. To maximize the signal/noise ratio, each point was obtained after a 5 second acquisition time and 5 spectra
    were measured for each temperature. The self-absorption correction and spectra normalization were applied as
     in Ref.~\onlinecite{Manuel2012}.

The temperature-dependent measurements of $^{25}$Mg and $^{17}$O static isotropic chemical shifts in MgO are
 taken from Ref.~\onlinecite{Fiske1994}. The NMR active isotopes $^{25}$Mg and $^{17}$O have  weak natural 
 abundances and gyromagnetic ratios, thus experiments required isotopically enriched samples. Both nuclei are 
 quadrupolar ($I_{\mathrm{spin}}=5/2$) but, as any atomic site in MgO presents  $O_h$ symmetry, the experimental 
 EFG vanishes. Therefore, NMR peaks do not suffer from quadrupolar broadening and shifting.

\subsection{Calculation details}

 \begin{table}[h]
\caption{\label{tab:pp} Parameters of generation for Troullier-Martin ultrasoft (US) and  norm-conserving (NC) 
pseudopotentials.  Bessel functions are used to pseudize the augmentation charges. The radii are in Bohr units.}
\centering
\begin{tabularx}{\linewidth}{lXl}
\hline\hline \\[-2.0ex]
Atom & Valence states (Radius)   & Local part \vspace{2pt}\\ 
\hline \\[-2.0ex]
Mg (NC) & 3$s^{1}$(2.00) 3$p^0$(2.00) 3$d^{0}$(2.00) & $d$ \\
O  (NC) & 2$s^2$(1.45) 2$p^3$(1.45)  & $p$  \vspace{2pt}\\	
\hline \\[-2.0ex]
Mg\footnotemark[1]\footnotetext{The pseudopotential of the absorbing Mg atom was generated using the same 
parameters but with only one $1s$ core electron and used in the XANES calculation. } 
(US) & 3$s^{2}$(2.50) 3$p^0$(2.60)  3$d^{0}$(2.30) & $d$\\
O  (US) & 2$s^2$(1.35) 2$p^4$(1.35)  & $p$  \vspace{2pt}\\	
\hline\hline
\end{tabularx}
\end{table}

All the calculations were performed using the pseudopotential, plane wave  Q\textsc{uantum} ESPRESSO suite of 
codes,\cite{Giannozzi2009} within the DFT-PBE generalized gradient approximation (GGA).\cite{Perdew1996} The 
details of the pseudopotential used herein are given in Table~\ref{tab:pp}. Most of the calculations were done using 
ultrasoft \cite{Laasonen1993} GIPAW\cite{Pickard2001} pseudopotentials except for the NMR calculations for which 
norm-conserving Troullier-Martin\cite{Troullier1991} GIPAW pseudopotentials were preferred.

The method detailed in Sec.~\ref{sec:theory} has been carried out  creating the configurations at the Quasi-Harmonic 
level with the Stochastic Self-Consistent Harmonic Approximation (SSCHA) code\cite{Errea2013,Errea2014}  starting 
from the QHA dynamical matrices of MgO, which crystallizes in  space group $Fm\overline{3}m$ with  a room temperature 
lattice parameter $a=\SI{4.21}{\angstrom}$.  The NMR and XANES calculations were performed for temperatures ranging 
from 0~K to 1273~K with the experimental lattice parameters,\cite{Hazen1976}  so that the configurations and spectroscopic 
results account for the thermal-expansion  anharmonic effect. The 0~K calculations were performed using the 12~K lattice
 parameter.\cite{Reeber1995}

Self-consistent electronic densities were calculated in the unit-cell at the volume of each corresponding temperature. A 
$4\times4\times4$ $\mathbf{k}$-point grid sampled the Brillouin zone,\cite{Monkhorst1976} and a plane-wave energy 
cutoff (density) for the wave functions of 45 (540)~Ry was chosen. Then, the dynamical matrices were calculated using 
Density-Functional Perturbation Theory\cite{Gonze1995,Baroni2001}  on a $\mathbf{q}$-point grid commensurable with 
the supercell size needed for NMR and XANES calculations thereafter. Long-range electrostatic interactions were taken 
into account by calculating Born effective charges and electronic dielectric tensor.\cite{Gonze1997}

NMR calculations were performed using the \texttt{GIPAW} package.\cite{Pickard2001,Yates2007} For each temperature, 
the calculations were conducted for  $2\times2\times2$ supercell configurations containing 64 atoms. Each calculation was
 done on a $2\times2\times2$ $\mathbf{k}$-point grid, with a plane-wave energy cutoff for the wave functions (resp. density)
  of 90 (resp. 360)~Ry. The convergence was reached for ten configurations at  each temperature. For a given temperature, 
  640 isotropic shielding tensors were calculated and averaged  for each nucleus.  In this work, $\sigma_{\mathrm{ref}}$ is
   chosen so that experimental and calculated values match at room temperature. The principal components $V_{xx}$, 
   $V_{yy}$ and $V_{zz}$ of the  EFG tensor defined with $\left|V_{zz}\right|>\left|V_{xx}\right|>\left|V_{yy}\right|$
    are obtained by diagonalization of the tensor. The quadrupolar  coupling constants $C_Q$  defined as $C_Q=eQV_{zz}/h$
     were estimated with  $Q$ values from Ref.~\onlinecite{Pyykko2008}  and were found  negligible  at each temperature. 
     This behavior is consistent with the crystal symmetry and experimental results. Indeed, in a  perfect cubic environment we expect the first-order EFG tensor quantities to vanish.

For each temperature, the XANES spectra were calculated  using  the \texttt{XSpectra} 
package\cite{Taillefumier2002,Gougoussis2009} in  $3\times3\times3$ supercell configurations containing 216 atoms, 
each including one 1$s$ full core-hole in a random Mg site.  The self-consistent electronic density was obtained at the
 $\Gamma$ point of the Brillouin zone with a plane-wave energy (resp. density) cutoff  of 45 (resp. 540)~Ry. The XANES 
 theoretical spectra were performed on a $4\times4\times4$ $\mathbf{k}$-point grid with a constant broadening parameter 
 of 0.5~eV.  For a given configuration,  three polarized XANES spectra were  calculated for the X-ray polarization vector
  $\mathbf{\hat{e}}$ parallel to each  of the Cartesian axes and the average gave an  isotropic XANES spectrum.    As a 
  convergence criterium, it has been verified that the averaged polarized spectra along each Cartesian direction matched, 
  as expected in a cubic crystal.  The convergence was reached for a number of 30 configurations at each temperature. In
   the pseudopotential approximation, only valence-electrons  are considered, hence the energy scale has no  physical 
   meaning and a core-level shift has to be applied before comparing and averaging  theoretical spectra (Eq.~\ref{eq:statavmc}). 
Similarly to previous works,\cite{Mizoguchi2009a,Jiang2013,Lelong2014} the core-level shift in the $i$th configuration is
 taken into account as follows:
\begin{eqnarray}
E \rightarrow E - \varepsilon_{\textsc{lub}}^i + \left( E_{\textsc{xch}}^i - E_{\textsc{gs}}^i\right).
\end{eqnarray}
In this rescaling, the  energy of the lowest unoccupied electronic band $\varepsilon_{\textsc{lub}}^i$ was subtracted 
and  the energy difference between the system with one $1s$ core-hole and one electron in the  first available 
electronic state ($E_\textsc{xch}^i)$ and that of the ground state ($E_\textsc{gs}^i$) was added. Finally,  all the 
spectra were shifted by 1303.8~eV to match   the experimental  main edge peak energy position at room temperature.
To interpret XANES spectra, local and partial density of states (DOS) were calculated in the $3\times3\times3$ supercell 
configurations using L\"owdin projections on a $4\times4\times4$ $\mathbf{k}$-point grid with a Gaussian broadening 
parameter of 0.3~eV.


\section{Results and discussion}
\label{sec:results}

\subsection{Solid-state NMR spectroscopy}

\begin{figure*}[]
\includegraphics[height=\linewidth,angle=270]{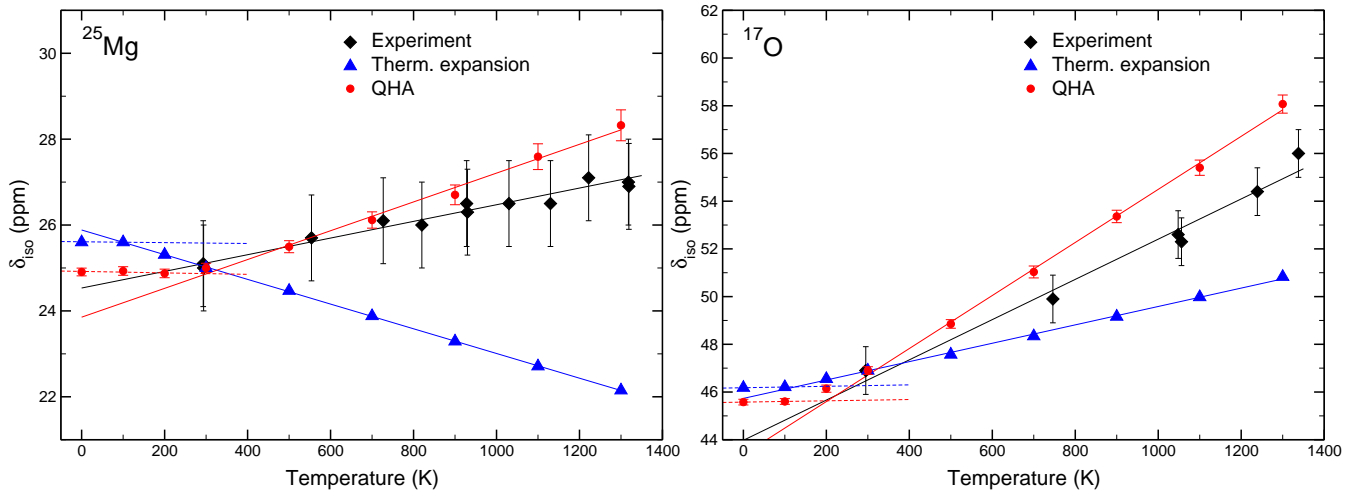}
\caption{\label{fig:nmr} (Color online)~Temperature-dependent isotropic chemical shift of $^{25}$Mg and $^{17}$O 
in MgO obtained experimentally\cite{Fiske1994} (diamonds),   calculated  in the QHA  (circles), and considering 
only the thermal expansion (triangles). The linear fits of each  dataset  are displayed. For the QHA calculation, 
two independent linear contributions are observed: from 0~K to 100~K and from 300~K to 1273~K. The 
experimental errors bars are set to $\pm 1$~ppm as in the original paper, while the calculated  error bars 
are the statistical uncertainties.}
\end{figure*}

Figure~\ref{fig:nmr} compares the temperature-dependence of the experimental NMR measurements from 
Ref.~\onlinecite{Fiske1994} with the calculated isotropic chemical shifts $\delta_{\mathrm{iso}}$ of $^{25}$Mg 
and $^{17}$O in MgO  up to 1273~K.  For both nuclei, the calculated values closely reproduce the 
experimental trend: the chemical shift increases with temperature. In addition, the contribution of the 
thermal expansion is displayed. The incorporation of the quantum motion of nuclei is mandatory, since 
it improves the agreement with experiment, especially for $^{25}$Mg. Without the quantum motion, 
considering only the thermal expansion leads to a downward trend in $^{25}$Mg and an upward one 
in $^{17}$O. These opposite trends  come from the fact that unlike $^{25}$Mg, the chemical shift of 
$^{17}$O is not a function of the Mg-O bond length. Indeed, the chemical shift of $^{17}$O depends 
on the interaction between the empty $d$ states of $^{25}$Mg and the $2p$ states of $^{17}$O, which 
results in a deshielding of $^{17}$O.\cite{Profeta2004}

\begin{table}[h]
\caption{\label{tab:slopesHT} Slopes of the temperature-dependent chemical shifts $\delta_{\mathrm{iso}}$
 for temperatures ranging from 300~K to 1300~K, in ppm.K$^{-1}$. The given values are the best fit lines in 
 the room-to-high temperature region. The experimental results are reported  and compared   to calculated 
 values from this work and  previous studies.}
\centering
\begin{tabularx}{\linewidth}{XXXXX}
\hline\hline \\[-2.0ex]
Nucleus &  Expt.\footnotemark[1]\footnotetext{Ref.~\onlinecite{Fiske1994}} & This work  &  Calc.\footnotemark[2]\footnotetext{Ref.~\onlinecite{Rossano2005}} & Calc.\footnotemark[3]\footnotetext{Ref.~\onlinecite{Monserrat2014}}\\
\hline \\[-2.0ex] 
$^{25}$Mg & 0.002 &  0.003 & 0.004 & 0.005  \\
$^{17}$O & 0.008 &  0.010 & 0.011 & 0.005  \\	
\hline\hline 
\end{tabularx}
\end{table}

The  linear trend behavior of the chemical shift observed from 300~K to 1273~K is consistent with experiments
 and previous theoretical studies.\cite{Rossano2005,Monserrat2014} Our results show  a  better agreement with
 experiments than other theoretical studies, the corresponding slopes obtained for each nucleus are
  summarized in Table~\ref{tab:slopesHT}.   The method used in Ref.~\onlinecite{Rossano2005}  is similar to 
  ours, but the remaining discrepancies  may come  from a less accurate description of the phonons dispersion 
  and quasi-harmonic vibrational wave functions. In Ref.~\onlinecite{Monserrat2014}, the coupling between 
  the phonons and the chemical shift tensor was expanded in terms of    vibrational-mode amplitudes and was
    assumed to be quadratic. However,  in this study, the   thermal expansion was neglected, thus, the results 
    cannot be compared to  experiment and the same slope is reported for the temperature-dependence of both 
    nuclei unlike experiments.

The linear behavior in the room-to-high temperature region  arises from a combination of both the thermal 
expansion and  dynamics of nuclei.  These two effects are constant in the low-temperature regime, from 0~K to 
100~K, where the vanishing thermal expansion  gives rise to a flat temperature dependence and the dynamics 
of the nuclei reduces to a $\Delta\delta_{\mathrm{iso}}$ constant term (Fig.~\ref{fig:nmr}). The zero-point 
renormalization of the chemical shift is evaluated to $\Delta\delta_{\mathrm{iso}}(^{25}\mathrm{Mg})=0.69$~ppm 
and $\Delta\delta_{\mathrm{iso}}(^{17}\mathrm{O})=0.60$~ppm, respectively. Above 300~K, the amplitude of the
 dynamical part outweighs the thermal expansion. Finally, in the intermediate temperature region (typically between
  100~K and  300~K) there is a competition between the two components of the temperature-dependence that leads 
  to the  observed curvature.

The origin of the remaining discrepancies between calculated and experimental slopes has to be investigated.
 The  overestimation of the temperature-dependence of both nuclei chemical shifts    could be related to a 
  deficiency of  GGA. Indeed, it is well known that  GGA  overestimates slightly the interatomic distances of 
  solids and underrates  phonon frequencies.\cite{He2014} Although  experimental lattice parameters are 
  chosen to balance this effect,  the use of more accurate GGA functionals\cite{Wu2006,Perdew2008} might 
  enhance the agreement. Furthermore,  it has been noticed that, in ionic compounds containing alkaline-earth
   cations,  the calculation-experiment  correlation lines of NMR parameters, as calculated within  GGA, deviate
   from the 1 ideal value.\cite{Laskowski2013}   In the case of CaO,  it has been observed that the shortening of 
   the  bond length induces an artificial hybridization between the  $3d$ states of Ca and the $2p$ of O.\cite{Profeta2004} 
   This problem was addressed by the use of a corrected Ca pseudopotential. To a lesser extent this problem can occur
    in MgO. \citet{Laskowski2013} proposed to use GGA$+U$ instead of standard GGA to better describe the $3d$
     states of the cation. Finally, one may argue of  a failure of  QHA, but the theory has revealed to accurately provide
      the lattice dynamical properties in MgO up to 1100~K under ambient pressure conditions.\cite{Wentzcovitch2010} 
      To go beyond  QHA, anharmonic effects in periodic solids can be investigated in a numerically feasible manner 
      using the vibrational self-consistent field method \cite{Monserrat2013} or the SSCHA.\cite{Errea2014}

\subsection{XANES spectroscopy}

\begin{figure}[]
\centering
      \includegraphics[width=\linewidth]{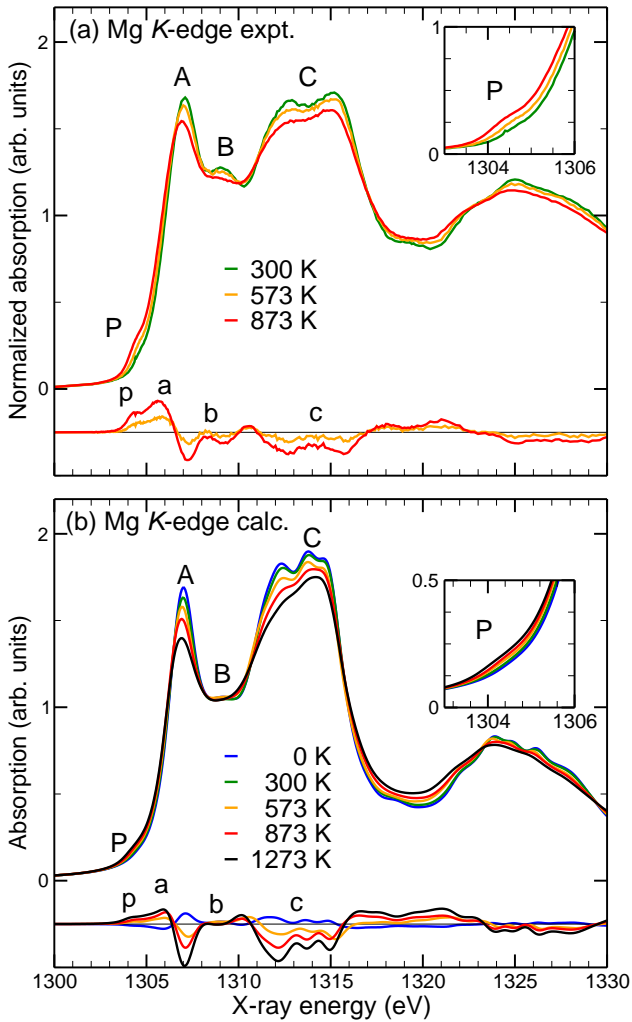}
\caption{(Color online)~Mg $K$-edge XANES experimental  (a)  and  QHA-calculated (b) spectra of MgO, for 
temperature up to 1273~K, along with difference of each spectrum with respect to the room-temperature one.}
\label{xanes.all}
\end{figure}

In Fig.~\ref{xanes.all}(a) the experimental Mg $K$-edge XANES spectra are reported. While  temperature 
continuously smoothes the XANES features, the P pre-edge peak  increases and shifts towards lower energy. 
To a lesser extent, this energy shift is also visible for the A main edge peak and the following features. These 
temperature effects have been also observed at the Al $K$-edge in  corundum ($\alpha$-Al$_2$O$_3$) and 
beryl (Be$_3$Al$_2$Si$_6$O$_{18}$).\cite{Manuel2012} The corresponding calculated  XANES spectra are 
plotted in Fig.~\ref{xanes.all}(b) along with the 0~K and 1273~K spectra. Calculations reproduce closely   
experiments over all the explored incident X-ray photon energy range, as highlighted by the similarity of the
 difference of each spectrum with the room-temperature one.  However, the intensity of the calculated pre-edge 
 is underestimated: the experimental and calculated P/A intensity ratios   are about 1/5 and 1/7, respectively.  
 This mismatch  may come from the first-order expansion of  the X-ray absorption cross section (Eq.~\ref{eq:avt}). 
 Going further than the first-order  is a possible improvement  of the method. In addition, as for NMR, 
 configurations obtained  beyond  QHA could lead to a better agreement with experiment.\cite{Monserrat2013,Errea2013}

\begin{figure}[]
      \includegraphics[width=\linewidth]{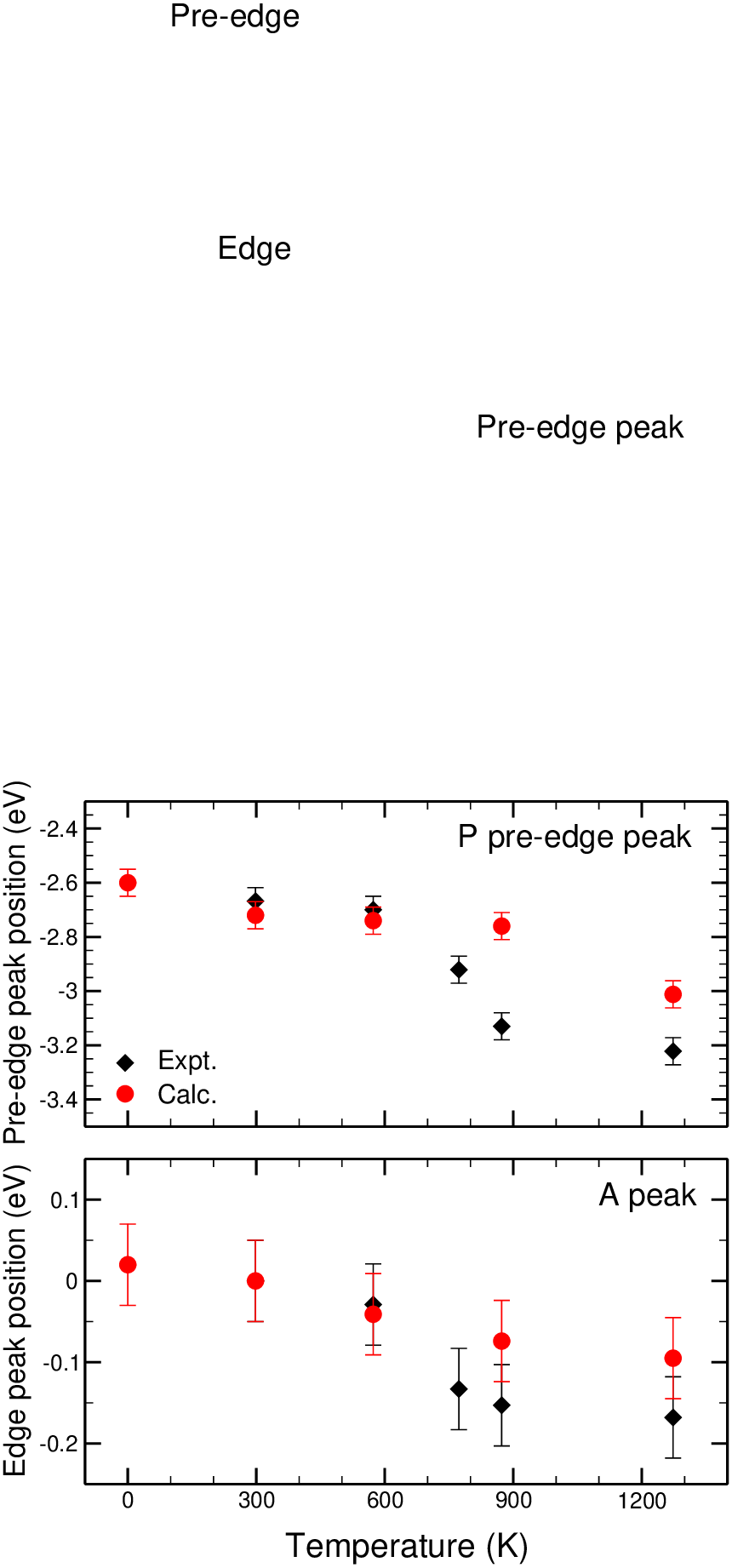}
\caption{\label{fig:peakvariation}
(Color online)~Experimental and calculated energy positions of P (upper panel) and A (lower panel) peaks as 
a function of temperature. The energy positions are given with respect  to the A peak position at 300~K.}
\end{figure}

The temperature-dependence of the P and A peaks energy positions is showed in  Fig.~\ref{fig:peakvariation}. 
For temperature ranging from 300~K to 1273~K,  the theory-experiment agreement is satisfactory:  the 
experimental and calculated P peak variation is 0.55~eV and 0.41~eV, respectively. For  peak A, the 
experimental (resp. calculated)  variation is   0.17~eV (resp. 0.12~eV).   The results are comforted by the 
experimental  observations at the Al $K$-edge in corundum where the pre-edge position decreased of about 
0.4~eV from 300~K to 930~K.\cite{Manuel2012}  The origin of these shifts may be related to the band gap 
evolution in temperature. In ionic compounds, the lattice expansion affects electronic bands. In MgO, the 
band gap was shown to decrease by 0.91~eV, from 300~K to 1273~K, using optical reflectivity 
measurements.\cite{Bortz1990} Moreover, the electron-phonon interaction contributes more than the  thermal 
expansion to the band gap narrowing.\cite{French1990,OBrien1992}  Over the same range of temperature, 
our calculations achieved a similar  band gap decrease ($0.78$~eV), while  the decrease only due to thermal 
expansion  is   $0.3$~eV.

\begin{figure}[]
\centering
\includegraphics[height=\linewidth,angle=270]{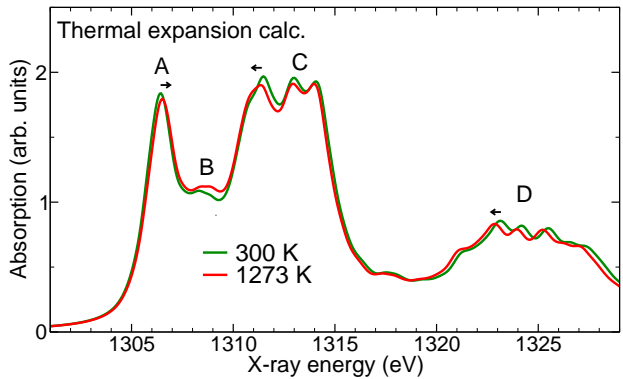}
\caption{(Color online)~Calculated Mg $K$-edge XANES spectra in MgO considering only the thermal expansion.}
\label{xanes.dilat}
\end{figure}

The energy shift to lower energy does not exclusively concern the P and A features since it is visible over 
all the spectral  energy range. For instance, peak D moves of about 0.1~eV to lower energy as temperature 
increases. The decreasing shift  agrees with the empirical Natoli's rule ($Ed^2=\mathrm{cste}$),\cite{Natoli1984} 
which states that the energy position decreases with increasing interatomic distance. Therefore,  this signature 
can be related to the thermal expansion. Figure~\ref{xanes.dilat} displays the Mg $K$-edge XANES spectra  
calculated in the equilibrium configuration at the  volumes corresponding to 300~K and 1273~K. The 1273~K 
spectrum is more contracted than the  300~K spectrum. The C and D features move down toward lower 
energy with increasing interatomic distance as observed experimentally. On the contrary, in opposite trend 
with experiment, peak  A  shifts toward  higher energy.   Therefore, thermal expansion does not fully explain 
the shifting trends, especially at lower energy, where vibrations are mandatory to reproduce the correct 
spectral-feature positions.

\begin{figure}[]
\centering
\includegraphics[height=\linewidth,angle=270]{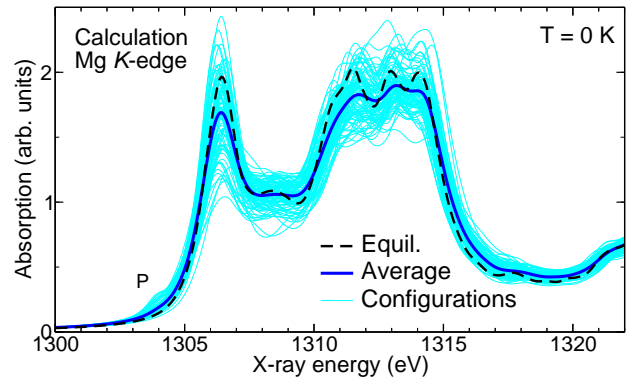}
\caption{(Color online)~Impact of the zero-point motion on the theoretical Mg $K$-edge XANES spectrum of MgO. The dashed spectrum is obtained with the atoms fixed at their equilibrium positions in the 12~K experimental volume.\cite{Reeber1995} The solid spectrum is the average of the 30 configuration spectra (in light blue).}
\label{xanes.300}
\end{figure}

Former studies already proposed that the P peak originates from the vibration-induced 
violation of the dipole forbidden $1s \rightarrow 3s$ transitions.\cite{Cabaret2009,Manuel2012,Brouder2010} 
To further analyze the P peak origin,  Fig.~\ref{xanes.300} compares the theoretical XANES 
spectra obtained at 0~K, by including or not the zero-point motion. In addition, 
Fig.~\ref{xanes.300} displays all the core-level shifted individual  configuration spectra  in the 
background. The phonon influence on the XANES spectra is characterized by two main features. 
First, similarly to a convolution and in agreement with the theoretical framework of  
Fujikawa,\cite{Fujikawa1999}  the inclusion of the 0~K quantum fluctuations globally smoothes 
the XANES spectrum.  Second, the zero-point motion induces a pre-edge peak that is totally absent 
in the equilibrium spectrum, i.e., when the atoms are fixed at their equilibrium positions. Hence,
 Fig.~\ref{xanes.300} highlights the quantum origin of the pre-edge. Moreover, the weak difference in 
 the pre-edge intensities between 0~K and 300~K calculations  emphasizes  the prominent role of 
 quantum effects up to  room temperature [Fig.~\ref{xanes.all}(b)].  A better description of the P and A 
 peaks intensities and variations could be achieved by the phonon-renormalization of the electronic energies
  directly in the self-consistent calculation, as performed in Refs.~\onlinecite{Antonius2014,Ponce2014a,Marini2015,Ponce2015}.

\begin{figure*}[]
   \begin{minipage}[c]{.32\linewidth}
      \includegraphics[width=\textwidth]{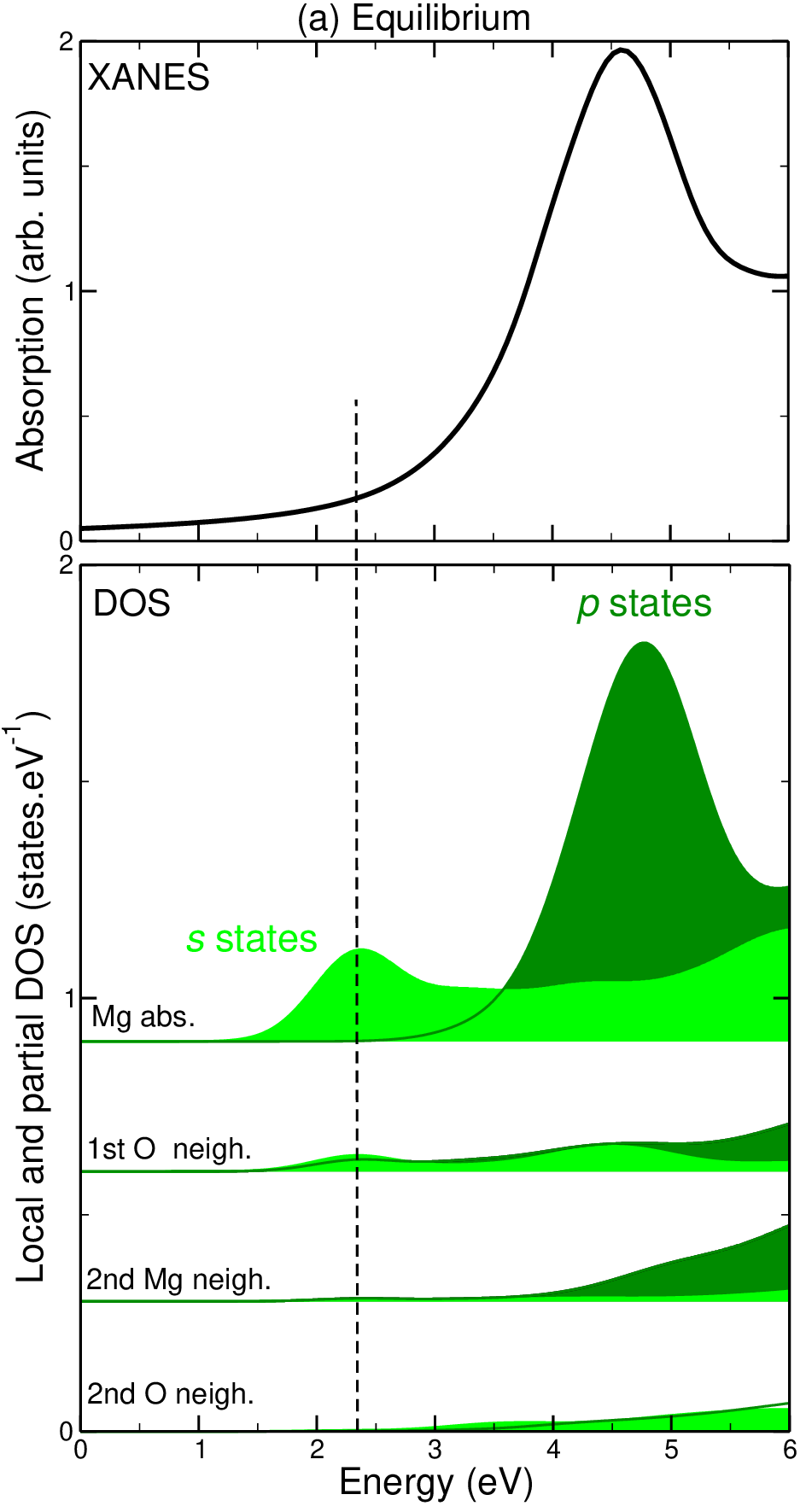}
   \end{minipage} \hfill
   \begin{minipage}[c]{.32\linewidth}
      \includegraphics[width=\textwidth]{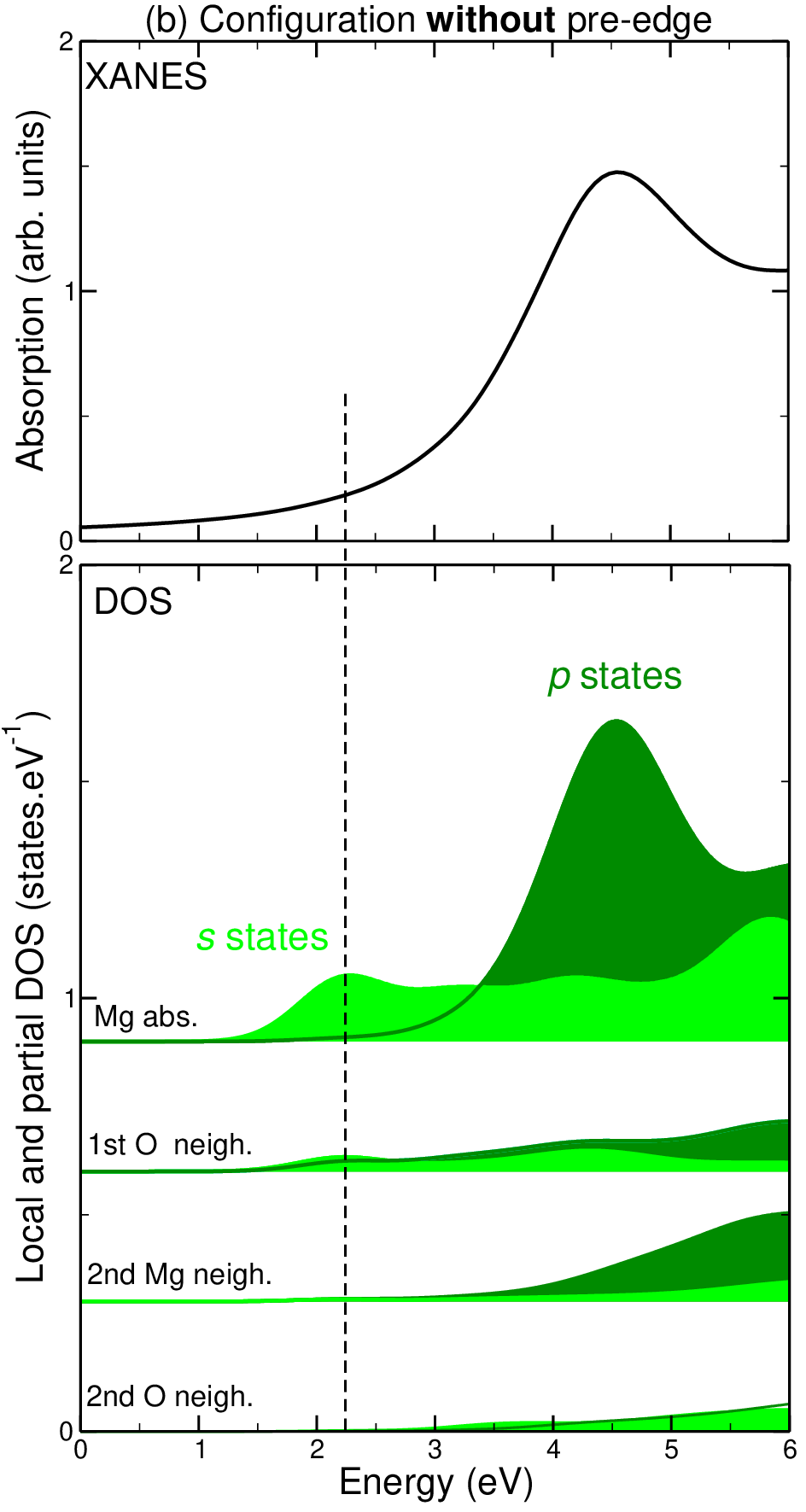}
   \end{minipage} \hfill
   \begin{minipage}[c]{.32\linewidth}
         \includegraphics[width=\textwidth]{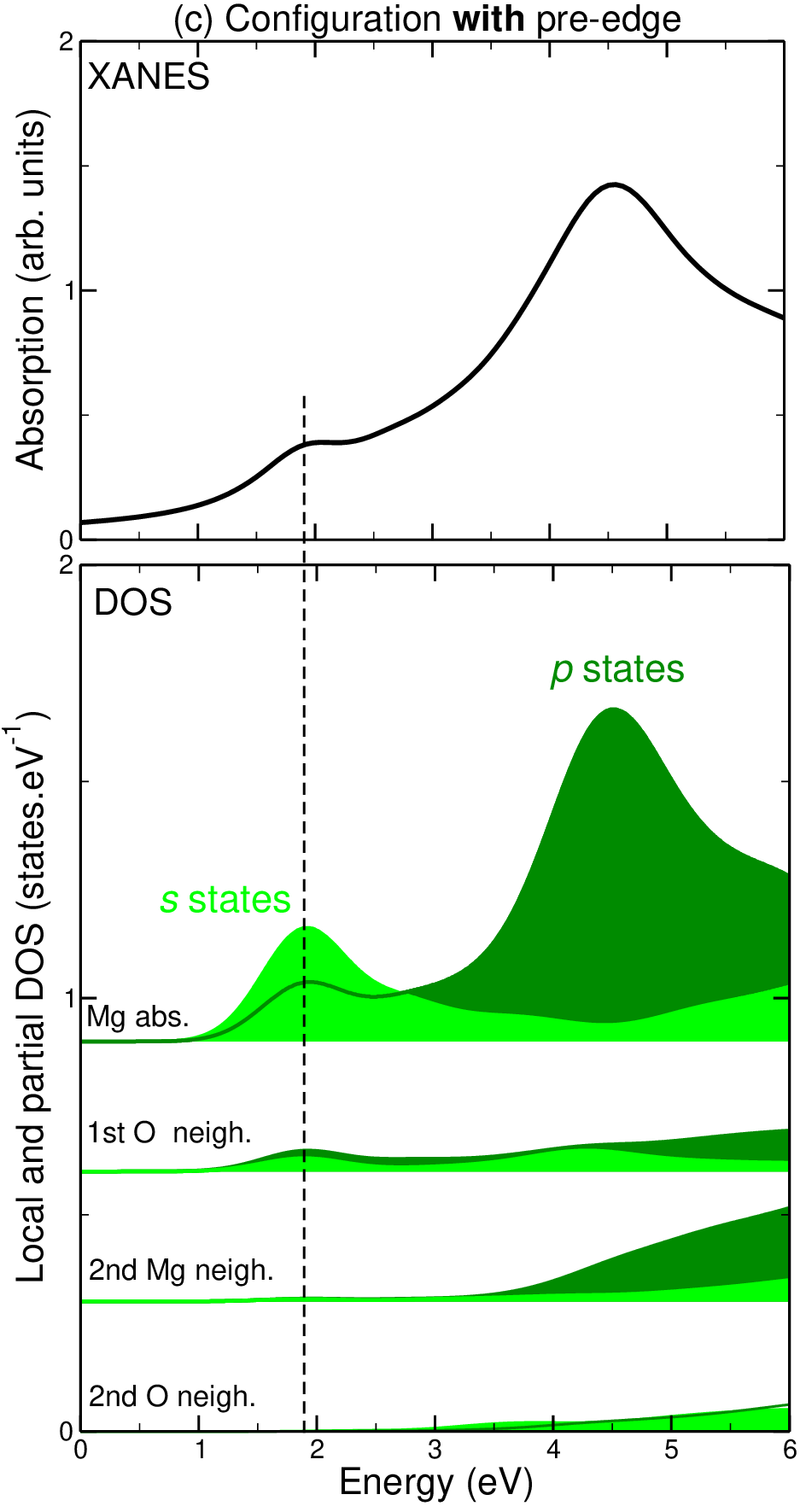}
   \end{minipage}
      \begin{minipage}[c]{\linewidth}
      \flushright
         \includegraphics[height=17.8cm,angle=270,trim=6cm 1.5cm 6cm 1.5cm]{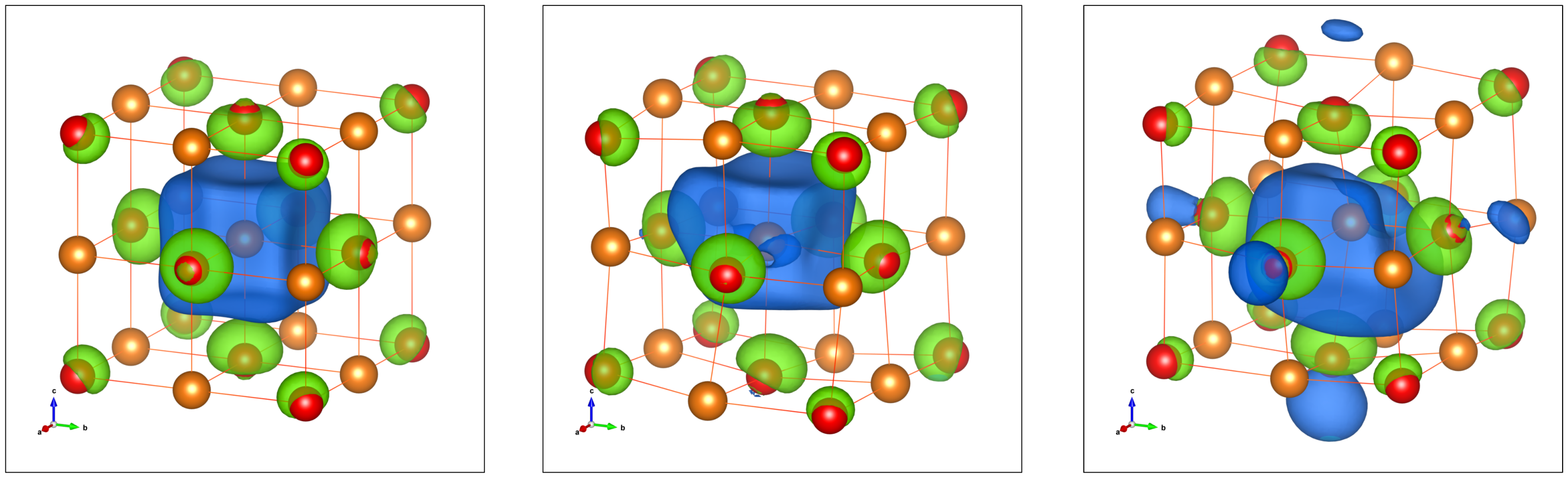}
                  \end{minipage}
   \caption{\label{fig:dos} 
   (Color online)~Pre-edge peak analysis. Three configurations at 1273~K are considered: the equilibrium structure (a), an individual configuration not leading to pre-edge peak (b) and an individual configuration that leads to pre-edge peak (c).   
    Top panel: Mg $K$-edge theoretical XANES spectra.   
    Middle panel:  local and partial electronic density of states (see text for details). 
    Bottom panel: isosurfaces of $\mathrm{sign}(\psi_{\textsc{lub}})\left| \psi_{\textsc{lub}} \right|^2$, where $\psi_{\textsc{lub}}$ is the electronic wave function of the lowest unoccupied band, whose energy is indicated by the vertical dashed line in top and middle panels (negative and positive signs are colored in green and  blue, respectively).  The Mg atoms are displayed in orange and O atoms in red. The  pictures are centered on the absorbing Mg atom.  The isosurface level is set to $6\times10^{-4}$~$a_0^{-3}$, with $a_0$ the Bohr radius.}
\end{figure*}

The 0~K individual configuration spectra   present a strong dispersion around  the averaged 
spectrum (Fig.~\ref{xanes.300}).  This  dispersion  increases with temperature. Some of the 
individual configuration spectra exhibit a strong P peak while some other do not, whatever 
the temperature.   Two individual configuration XANES spectra at $T=1273$~K, one with a 
pre-edge and one without, have been selected  and analyzed using   local and partial DOS 
[Figs.~\ref{fig:dos}(b,c)]. For comparison, Fig.~\ref{fig:dos}(a)  displays the case of the standard 
calculation with atoms fixed at their equilibrium positions at $T=1273$~K. The partial and local 
DOS plotted in Fig.~\ref{fig:dos} are the   $s$ and $p$ empty DOS projected on (i) the absorbing 
Mg, (ii) the first   O neighbors, (iii) the next Mg neighbors, and (iv) the next O neighbors. 
Whatever the configuration, the core-hole strongly modifies the $s$ and $p$ empty states of Mg, 
leading to two rather localized peaks, which coincide with P and A, respectively. P is only visible 
if $sp$ hybridization of the Mg absorbing states  occurs [Figs.~\ref{fig:dos}(c)]. This $sp$ mixing 
of the Mg absorbing states also induces a stronger $p$ DOS of oxygen. The  $sp$ hybridization 
due to the  dynamical distortion of the MgO$_6$ octahedron  does not systematically occur 
[Fig.~\ref{fig:dos}(b)]. The contribution of the lowest unoccupied band  to the   electronic charge 
density   ($\pm\left| \psi_{\textsc{lub}} \right|^2$) gives a representation of the electronic state 
probed in the P peak and enlightens the effect of the MgO$_6$ octahedron distortion.   In the 
equilibrium configuration [Fig.~\ref{fig:dos}(a)] the  isosurface shows a centrosymmetric cubic 
shape.   The distortion of the lattice strongly impacts $\pm\left| \psi_{\textsc{lub}} \right|^2$ 
[Fig.~\ref{fig:dos}(b)], however, it is not sufficient to create a pre-edge peak. The P peak emerges 
if  the distortion of the MgO$_6$ octahedron  induces  a $p$-like character on the neighboring 
O atoms, as already observed in the DOS [Fig.~\ref{fig:dos}(c)]. To conclude, the breakdown of 
the symmetry is mandatory to induce the $sp$ hybridization and the forbidden 
$1s \rightarrow 3s$ transition but  is not sufficient.

\section{Conclusion}
\label{sec:end}

A DFT-based approach  enabling to successfully introduce quasi-harmonic  quantum 
thermal fluctuations of nuclei in NMR and XANES spectroscopies has been presented. 
This method, avoiding the explicit calculation of the electron-phonon coupling parameters, 
provides an efficient framework to analyze phonon effects occurring in both spectroscopies. 
The calculated spectral data obtained in the MgO proof-of-principle compound are in good 
agreement with experimental datasets, which  supports the reliability of our approach. 

The combination of experiments and first-principle calculations have enabled to 
investigate the influence of the quantum vibrations  in both spectroscopies. A similar 
behavior is revealed: the zero-point phonon renormalization of NMR and XANES spectra 
improves the experiment-calculation agreement and therefore could be used on a regular 
basis to reproduce experimental data even at low temperatures. 

In  NMR, the experiment-calculation agreement is improved with respect to previous 
theoretical studies. The temperature-dependence of the chemical shifts results from 
both contributions of thermal expansion and nuclear dynamics, and   reduces to a 
constant renormalization term at low temperature.

In the case of XANES, the  temperature-dependence of XANES features is reproduced 
over a large range of temperatures. An analytic expression of the phonon-dependent X-ray
 absorption cross section is derived at the first  order of the electronic Green's function 
 expansion. It appears that keeping only the first term in the expansion of Eq.~(\ref{gtg}) 
 is a suitable approximation to calculate XANES spectra at finite temperature. It appears 
 that the first-order calculation of  XANES spectra at finite temperature is a suitable 
 approximation. Nevertheless, the implementation of the higher-orders  correction terms 
 could improve the pre-edge intensity modeling. The presence of the pre-edge feature is a 
 relevant signature of phonon effects. A thorough study of the mechanism from which the 
 pre-edge emerges is conducted. The breakdown of the coordination symmetry is mandatory 
 to induce the pre-edge and a $p$-like character arises on the neighboring O atoms if the 
 pre-edge feature is discernible.  The pre-edge energy variation in temperature is related to 
 the band gap temperature-dependence, whereas the variation of the high-energy structures 
 originates from the thermal expansion.

The results obtained for MgO can be extrapolated to other light-element oxides. In 
the case of XANES spectroscopy, the method is applied to corundum in a forthcoming 
publication.\cite{nemausat2015d} Techniques closely related to XANES, such as Non 
Resonant Inelastic X-ray Scattering and core-loss Electron Energy Loss Spectroscopy, 
may also be affected by vibrations and could highly benefit from this theoretical framework.

\begin{acknowledgments}

This work was supported by French state funds managed by the ANR within the 
Investissements d'Avenir programme under reference  ANR-11-IDEX-0004-02, and 
more specifically within the framework of the Cluster of Excellence MATISSE led by 
Sorbonne Universit\'es. It was performed using the HPC resources of GENCI-IDRIS 
(grants: 2015-100172). Experiments were performed on the LUCIA beam line at 
SOLEIL Synchrotron (proposal: 20141057). The authors acknowledge Guillaume Radtke,  
\'Etienne Balan, Delphine Vantelon, Jean-Paul Iti\'e and Yves Joly for very fruitful 
discussions and Lorenzo Paulatto for his help with the SSCHA code.
\end{acknowledgments}

\appendix

\section{Phonon-renormalized  X-ray absorption cross-section}
\label{green}

In this section we discuss two questions that are rarely addressed in the
literature. 
To describe the first question recall that, in optical spectroscopy of 
molecules, the equilibrium nuclear positions of the initial and final states of the
absorption process are different. This leads to 
vibronic effects that are often calculated by using the Franck-Condon principle.
In X-ray absorption spectroscopy, the experimental spectrum is well reproduced
although the final states are calculated for the same nuclear positions 
as the initial state. How can this be?

The second question has to do with the effect of temperature on 
X-ray absorption spectra. This effect is often calculated by
averaging spectra calculated over a distribution of atomic positions
corresponding to the temperature.\cite{Pascal2014,Pascal2015,Peyrusse2010,Mazevet2014}
In principle, this procedure
is not correct although it gives reasonable results in practice.
How can this be?

In the present section, we answer these two questions by showing
that the usual calculation methods amount to neglecting
the nuclear kinetic energy in the
absorption process and we show how
it is possible to go beyond this approximation.

To describe the effect of nuclear vibrations
on XANES spectra, we write the electric dipole
absorption cross-section in terms of the wave functions
$\ket{\Psi_n^j}$ involving both the electronic and
the nuclear variables:
\begin{eqnarray}
\label{eq:sigmatot}
\sigma_{tot}(\hbar\omega) &=& 4\pi^2\alpha_0 \hbar\omega \sum_{n,j} \left| \bra{\Psi_n^j} \mathcal{T} 
\ket{\Psi_0^0}\right|^2\delta(E_n^j - E_0^0 - \hbar\omega)
\nonumber\\&=& -4\pi\alpha_0 \hbar\omega
  \bra{\Psi_0^0}\mathcal{T}^\dagger \mathrm{Im} G(E_0^0+\hbar\omega) \mathcal{T}\ket{\Psi_0^0},
\end{eqnarray}
where we used:\cite{Newton1982}
\begin{equation*}
\sum_{n} \ket{\Psi_n^j}\delta(E_n^j - E_0^0 - \hbar\omega) \bra{\Psi_n^j} 
= -\frac{1}{\pi}\ \mathrm{Im}\left[G(E_0^0 + \hbar\omega)\right].
\end{equation*}
Equation~(\ref{eq:sigmatot}) is equivalent to Eq.~(\ref{eqxas2})
for the transition operator $\mathcal{T}=\mathbf{\hat{e}\cdot\overline{r}}=
\sum_i^{N_e} \mathbf{\hat{e}\cdot r}_i$.
The Green function $G$ of the full electronic and nuclear Hamiltonian 
is the solution of the following equation
\begin{equation}
\label{eqg0}
\left[z - H_{BO} - T_N\right] G(z) = 1,
\end{equation}
with the complex energy $z$ = $E$ + $\mathrm{i}\gamma$, where $\gamma$ an infinitesimal positive number 
or a finite number representing the broadening due to core-hole lifetime and experimental resolution. 
A straightforward expansion in Eq.~(\ref{eqg0}) gives
\begin{equation}
\label{gtg}G(z)=G_0(z)+G_0(z)T_NG(z),
\end{equation}
where 
$G_0(z) = \left[z - H_{BO}\right]^{-1}$
 is the Green function in the Born-Oppenheimer approximation.
 Equation~(\ref{gtg}) can be expanded into
\begin{equation}
G(z)=G_0(z)+G_0(z)T_NG_0(z) + \ldots\ .
\end{equation}
The resolution of X-ray absorption spectra is determined by the
lifetime of the core hole (0.36~eV for the $K$-edge
of Mg~\cite{Krause1979}). It may be considered that
the core hole lifetime will smooth out effects that involve
a much smaller energy. Since the 
nuclear kinetic energy, evaluated using the zero-point 
energy values,\cite{Irikura2007} is 0.02~eV,
it seems reasonable to neglect the kinetic energy 
$T_N$ and to keep only the first term of the expansion:   
\begin{equation}
G(z) \approx G_0(z). 
\label{ghg}
\end{equation}
The second term is expected to play a role in the presence
of forbidden transitions, but this effect will not be
considered in the present paper.
We reach the approximate absorption cross-section
\begin{equation}
\sigma_{tot}(\hbar\omega) = -4\pi\alpha_0 \hbar\omega\ 
\mathrm{Im} \bra{\Psi_0^{0}} \mathcal{T}^\dagger~G_0~
\mathcal{T}\ket{\Psi_0^{0}}.
\label{eq.exact}
\end{equation}
In this approximation, the calculation of the phonon-renormalized XANES 
cross section requires only the Born-Oppenheimer Green function $G_0$,
for which we now give the following convenient expression:
\begin{equation}
\bra{\mathbf{\overline{r}',\overline{R}'}}  G_0 
\ket{\mathbf{\overline{r},\overline{R}}}  = 
\sum_{n,j}\frac{\braket{\mathbf{\overline{r}',\overline{R}'}|\Psi_n^j}
\braket{\Psi_n^j|\mathbf{\overline{r},\overline{R}}}}{z-\varepsilon_n(\mathbf{\overline{R})}}.
\end{equation} 
The validity of this expression can be established by showing that
it solves the equation for $G_0$:
\begin{equation}
\left[z - H_{BO}\right]\ \bra{\mathbf{\overline{r}',\overline{R}'}} G_0 \ket{\mathbf{\overline{r},\overline{R}}} = \delta(\mathbf{\overline{R}'}-\mathbf{\overline{R}}) \ \delta(\mathbf{\overline{r}'}-\mathbf{\overline{r}}).
\end{equation}

Equation~(\ref{eq.exact}) is now evaluated as follows
\begin{multline}
\sigma_{tot}(\hbar\omega) = -4\pi\alpha_0 \hbar\omega   \int  d\mathbf{\overline{r}} d\mathbf{\overline{r}'} d\mathbf{\overline{R}} d\mathbf{\overline{R'}}\ \mathrm{Im} \bra{\Psi_0^0} \mathcal{T}^\dagger \ket{\mathbf{\overline{r}',\overline{R}'}} \\ \times  \bra{\mathbf{\overline{r}',\overline{R}'}}  G_0 \ket{\mathbf{\overline{r},\overline{R}}} \bra{\mathbf{\overline{r},\overline{R}}} \mathcal{T} \ket{\Psi_0^0}.
\end{multline}
This expression can be simplified by noticing that, in the
Born-Oppenheimer approximation, $G_0$ is diagonal
in the nuclear variables. Indeed, by writing the
initial and final states in the Born-Oppenheimer approximation (Eq.~\ref{eq:bowfc})
we obtain
\begin{multline}
\label{A11}
\bra{\mathbf{\overline{r}',\overline{R}'}} G_0 \ket{\mathbf{\overline{r},\overline{R}}} = \sum_{n} \frac{\psi_n^{*}(\mathbf{\overline{r}';\overline{R}'})\psi_n(\mathbf{\overline{r};\overline{R}})}{  z - \varepsilon_n(\mathbf{\overline{R}})}  \\ \times  
   \sum_{j} \chi_n^{j*}(\mathbf{\overline{R}'}) \chi_n^{j}(\mathbf{\overline{R}}). 
\end{multline} 
The completeness relation of Eq.~(\ref{eq.chi}) turns Eq.~(\ref{A11}) into
\begin{eqnarray}
\bra{\mathbf{\overline{r}',\overline{R}'}} g_0 \ket{\mathbf{\overline{r},\overline{R}}} &=& \delta(\mathbf{\overline{R} - \overline{R}}') \sum_{n}  \frac{\psi_n^{*}(\mathbf{\overline{r}';\overline{R}})\psi_n(\mathbf{\overline{r};\overline{R}})}{  z - \varepsilon_n(\mathbf{\overline{R}})},\nonumber\\
\label{A12} &=& \delta(\mathbf{\overline{R} - \overline{R}}')\ 
  \bra{\mathbf{\overline{r}'}}g_0(\mathbf{\overline{R}})\ket{\mathbf{\overline{r}}},
\end{eqnarray}
where $g_0(\mathbf{\overline{R}})$ is the electronic Green function 
for a system where the nuclei are fixed at position $\mathbf{\overline{R}}$.
In other words, $g_0(\mathbf{\overline{R}})$ is the solution of
\begin{equation}
\left(z - H_{BO}\right)\ \bra{\mathbf{\overline{r}'}} g_0(\mathbf{\overline{R}}) 
\ket{\mathbf{\overline{r}}} = \delta(\mathbf{\overline{r}'}-\mathbf{\overline{r}}),
\end{equation}
where $H_{BO}$ is evaluated at the nuclear positions
$\mathbf{\overline{R}}$.
Introducing Eq.~(\ref{A12}) in Eq.~(\ref{eq.exact})  the absorption cross-section at the 
first-order in $G_0$ is obtained:
\begin{multline}
\label{A13}
\sigma_{tot}(\hbar\omega) = -4\pi\alpha_0 \hbar\omega   
\int  d\mathbf{\overline{r}} d\mathbf{\overline{r}'} d\mathbf{\overline{R}} 
\ \mathrm{Im} \bra{\Psi_0^0} \mathcal{T}^\dagger 
\ket{\mathbf{\overline{r}',\overline{R}}} \\ \times  
\bra{\mathbf{\overline{r}'}} g_0(\mathbf{\overline{R}}) 
\ket{\mathbf{\overline{r}}} \bra{\mathbf{\overline{r},\overline{R}}} 
\mathcal{T} \ket{\Psi_0^0}.
\end{multline}
The result of Eq.~(\ref{A13}) implies that the XANES calculation requires only 
the energy surface --  nuclear configuration -- of the initial state. Hence, Eq.~(\ref{A13})  
justifies the use of the ground-state crystallographic structure in the initial (without core-hole) and final (with core-hole) states when calculating the XANES cross-section. The total wave functions can be expressed in the BO approximation (Eq.~\ref{eq:bowfc}) 
\begin{multline}
\sigma_{tot}(\hbar\omega) = -4\pi\alpha_0 \hbar\omega \ \mathrm{Im}  
\int d\mathbf{\overline{R}}\    \left| \chi_0^0(\mathbf{\overline{R}})\right|^2
     \\    \times \int d\mathbf{\overline{r}}d\mathbf{\overline{r}'}\  
\psi_0(\mathbf{\overline{r}',\overline{R}})\ \mathbf{\hat{e}}^* \cdot 
\mathbf{\overline{r}'}\  g_0(\mathbf{\overline{R}}) \ 
\mathbf{\hat{e}} \cdot \mathbf{\overline{r}}\ \psi_0(\mathbf{\overline{r},\overline{R}}).
\label{eq:avgmulti}
\end{multline}
Equation~(\ref{eq:avgmulti}) proves that the effect of thermal 
vibrations on XANES spectra can be obtained by averaging
individual XANES spectra for nuclear positions $\overline{R}$
weighted by the distribution function
$\left| \chi_0^0(\mathbf{\overline{R}})\right|^2$ 
computed from the ground vibrational mode in the ground state.

However, Eq.~(\ref{eq:avgmulti}) is expressed in a 
many-body framework, whereas $K$-edge XANES spectra 
are usually calculated in a single-electron framework. 
Since this reduction is a classical problem, we 
just give a sketch of the derivation.
If we rewrite Eq.~(\ref{eq:avgmulti}) in terms
of wave functions, we have to deal with matrix
elements such as
$\bra{\psi_n} \mathbf{\hat{e}} \cdot \mathbf{\overline{r}} \ket{\psi_0}$,
where $\ket{\psi_0}$ and $\ket{\psi_n}$ are $N_e$-body
wave functions. If we assume that these wave functions
can be expressed as Slater determinants, the fact that
$ \mathbf{\hat{e}} \cdot \mathbf{\overline{r}} $ is a single-body
transition operator implies:\cite{Cowan1981} 
\begin{equation}
\bra{\psi_n} \mathbf{\hat{e}} \cdot \mathbf{\overline{r}} \ket{\psi_0}
= 
  \int d\mathbf{r}\
  \phi_{\beta'}^*(\mathbf{r})\ \mathbf{\hat{e}} \cdot \mathbf{r}\
  \phi_\beta(\mathbf{r}),
\end{equation}
where $(\phi_\beta,\phi_{\beta'})$ is the only pair of one-electron
orbitals that are different in $\ket{\psi_0}$
and $\ket{\psi_n}$, where $\phi_\beta$ is occupied
in $\ket{\psi_0}$ and $\phi_{\beta'}$ in $\ket{\psi_n}$.
For a $K$-edge the resulting expression is

\begin{multline}
\sigma_{tot}(\hbar\omega) = -4\pi\alpha_0 \hbar\omega \ 
\mathrm{Im}  \int d\mathbf{\overline{R}}\    \left| 
\chi_0^0(\mathbf{\overline{R}})\right|^2     \\    
\times \int d\mathbf{r}d\mathbf{r'}\  \phi_{1s}(\mathbf{r',\overline{R}})\ 
\mathbf{\hat{e}}^* \cdot \mathbf{r'}\ g(\mathbf{r}',\mathbf{r};\mathbf{\overline{R}}) 
\ \mathbf{\hat{e}} \cdot \mathbf{r}\ \phi_{1s}(\mathbf{r',\overline{R}}),
\end{multline}
where
$g(\mathbf{r}',\mathbf{r};\mathbf{\overline{R}})
=\sum_{\beta'} \phi_{\beta'}^*(\mathbf{r}';\mathbf{\overline{R}})
\phi_{\beta'}(\mathbf{r};\mathbf{\overline{R}})/(z-e_{\beta'})$.
A similar expression can be obtained from more sophisticated
many-body perturbation theory.

Considering  the cross section in a given nuclear configuration 
from  Eq.~(\ref{eqxas}) gives
\begin{equation}
\sigma_{tot}(\hbar\omega) =   \int d\mathbf{\overline{R}}\    
\left| \chi_0^0(\mathbf{\overline{R}})\right|^2    
\sigma(\hbar\omega;\mathbf{\overline{R}})
\label{eq.xasmoy}
\end{equation}
and  we demonstrate, restricting ourselves to the first order in 
the expansion of $G(z)$, that to account for the nuclear motion 
in the XANES cross section one must average the individual 
configuration spectra using a probability distribution, which is 
consistent with  Eq.~(\ref{eq:statavmc}). We used a ground-state 
phonon wave function  $\chi_0^0(\mathbf{\overline{R}})$ for 
notational convenience. The generalization to a 
Boltzmann distribution $\mathbf{\rho(\overline{R}})$ 
of phonon states at finite temperature is straightforward 
and amounts to replacing $\left| \chi_0^0(\mathbf{\overline{R}})\right|^2$ 
by $\mathbf{\rho(\overline{R}})$ in Eq.~(\ref{eq.xasmoy}).

\end{document}